\documentclass[reprint, 10pt,superscriptaddress, aps, prl]{revtex4-2}
\usepackage{graphicx}
\begin{document}
	\title{Experimental Limits on Planetary Mass Primordial Black Hole Mergers}
	
	\author{William M. Campbell}\affiliation{Quantum Technologies and Dark Matter Labs, Department of Physics, University of Western Australia, 35 Stirling Highway, Crawley, WA 6009, Australia.}
	\email{william.campbell@uwa.edu.au}
	\author{Leonardo Mariani}\affiliation{Department of Physics "Giuseppe Occhialini",  University of Milano-Bicocca, Piazza dell'Ateneo Nuovo, 1 - 20126, Milan (Mi), Italy}
	\author{Michael E. Tobar}\affiliation{Quantum Technologies and Dark Matter Labs, Department of Physics, University of Western Australia, 35 Stirling Highway, Crawley, WA 6009, Australia.}
	\author{Maxim Goryachev}\affiliation{Quantum Technologies and Dark Matter Labs, Department of Physics, University of Western Australia, 35 Stirling Highway, Crawley, WA 6009, Australia.}
	\email{maxim.goryachev@uwa.edu.au}

	\begin{abstract}
		The multi-mode acoustic gravitational wave experiment (MAGE) is a high-frequency gravitational wave detection experiment featuring cryogenic quartz bulk acoustic wave resonators operating as sensitive strain antennas in the MHz regime. After 61 days of non-continuous data collection, we present bounds on the observable merger rate density of primordial black hole binary systems of chirp mass $1.2\times10^{-4}M_\odot<\mathcal{M}<1.7\times10^{-9}M_\odot$. The maximum achieved limit on the merger rate density is $\mathcal{R}<1.3\times10^{18}~\mathrm{kpc}^{-3}\mathrm{yr}^{-1}$, which corresponds to constraining yearly mergers to a distance of reach on the order of the solar system, or $1.0\times10^{-6}$ kpc during the observational period. In addition, we exclude significantly rare and strong events similar to those observed in previous predecessor experiments as non-gravitational background signals, utilising coincident analysis between multiple detectors.
	\end{abstract}
	\maketitle
	The detection and subsequent study of gravitational wave emission signatures sourced by binary systems of black holes provides a valuable probe into a wealth of physics due to their extreme nature. Whilst large mass black hole systems have been successfully detected \cite{abbott2016}, smaller sub-solar mass black hole binary systems are yet to be confirmed. A critical point of difference between the two mass regimes is that the production mechanism for sub-solar mass black holes confines them to be of primordial origin, given that large density perturbations in the early universe could give rise to the appropriate energy densities required. However, primordial mass black holes (PBH) \cite{Khlopov2010, Khlopov2024} may make up a significant fraction of the total mass density of dark matter \cite{Belotsky2014}. This motivates the search for their direct and indirect signatures through experiment, as well as the application of theoretical bounds.
	
	The search for high-frequency gravitational waves (HFGWs) at $f > 10$ kHz in general has gained recent interest \cite{Aggarwal2021, Aggarwal2025}. Many theories have hypothesized sources of transient HFGWs due to the collision of compact objects such as PBHs, stochastic background distributions that provide windows into early universe dynamics, and also coherent monochromatic sources due to processes involving exotic particle species. Multiple experimental proposals have thus been put forth for HFGW detection. Varying in technological composition and scope, such experiments are proposed to be sensitive to gravitational signatures across frequency regimes from kHz to GHz.  Meter-scale optical interferometers have already attained competitive sensitivity to stochastic background sources in the 1-100 MHz region \cite{Chou2017, Patra2024}.  Optically levitated sensor detectors have been proposed to exploit the coupling between gravitational waves and the acoustic modes of a suspended nanoparticle. These devices show promise with broadband sensitivity in the  1-300 kHz, \cite{Arvanitaki2013, Aggarwal2020}. Electromagnetic resonant cavities primarily utilized for pseudo-scalar dark matter detection can be made sensitive to HFGWs in narrow bands corresponding to the microwave frequency of the cavity \cite{Tobar2022}. In addition, other resonant mass type detectors have been proposed considering the deformation of electromagnetic cavities \cite{Berlin2023} as well as magnetic structures \cite{Domcke2024}.
	
	Due to the length and energy scales involved, HFGW detection is extremely challenging. Most proposed sources give rise to signals with reduced amplitude at higher frequencies; a natural result when considering the energy density of higher frequency gravitational radiation. In addition, observable PBH signatures face further challenges when considering the signal duration. The rapid in-spiral phase of light PBH mergers occurs over extreme time scales, meaning the signal may only be observable in a detector's bandwidth for an incredibly brief time interval. This can ultimately hinder efforts to detect PBH mergers with resonant systems. To combat such challenges, extremely high quality factors can boost on-resonance sensitivity and extend the detector response time. At the same time, multiple higher-order overtone modes in the same acoustic system can be leveraged to gain further bandwidth.
	
	The Multi-mode acoustic gravitational wave experiment (MAGE) \cite{Campbell2023} is a resonant mass HFGW detector operating in the 5-15 MHz region. Utilizing two extremely low-loss cryogenic quartz bulk acoustic wave resonators \cite{Galliou2013, Goryachev2014}, this detector displays strong sensitivity to external strain fields in multiple narrow frequency bands corresponding to the overtone modes of the bulk crystal medium. In an initial path-finding experiment based on the same technology, significantly rare and high-energy signals were observed \cite{Goryachev2021}. While it is generally accepted that the sources of these excitations are far too strong to be any sort of gravitational event \cite{Lasky2021, Guillem2021}, further iterations on the experimental design were required to confidently exclude such signals.
	
	In this work, we present the results of a first observational run of the MAGE experiment, in which no signatures of HFGWs were observed over a data collection period of 61 days. We utilise an optimal filtering approach to identify signals coincident with the two MAGE detectors, removing non-gravitational background signals by coincident analysis. The resulting observations allow for an exclusion bound to be placed on the merger rate density of planetary mass PBH binary mergers $\mathcal{R}$ which has units of events per kpc$^3$ per year.
	
\begin{figure}
		\centering
		\includegraphics[width=0.47\textwidth]{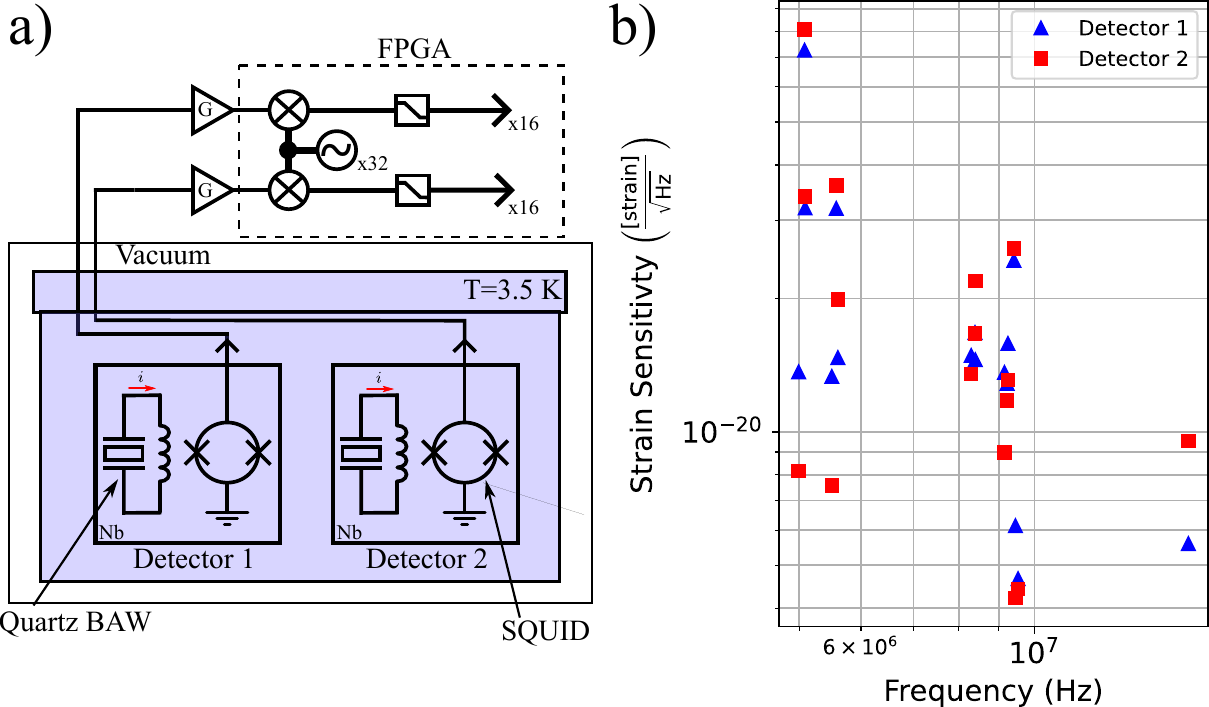}
		\caption{\label{fig:setup}a) Schematic of experimental setup for MAGE featuring the cryogenic system and data acquisition chain. b) Experimental strain sensitivity of each resonant mode used in this work.}
	\end{figure}

	Previous work has described the experimental setup of the MAGE  system in detail \cite{Campbell2023, Goryachev2014, Goryachev2014b}. In summary, the system features two near-identical HFGW detectors, each consisting of a piezoelectric quartz bulk acoustic wave resonator coupled to the input loop of a superconducting quantum interference device (SQUID). This allows for the piezoelectric charge distribution due to incident strains upon the crystal bulk to be amplified and readout as a voltage on the SQUID output. Each detector is hosted in an independent superconducting Niobium enclosure with various stages of electromagnetic shielding. The SQUID output voltage signal is sampled by an FPGA digitizer after additional amplification at room temperature. Re-configurable programming of the FPGA allows for the simultaneous and independent lock-in amplification of 16 different overtone modes in each detector, giving continuous monitoring of strain in multiple narrow frequency bands. Fig. \ref{fig:setup} a) gives a diagrammatic view of the experimental configuration.
    
	In a first observational run, MAGE collected approximately two months of data from November 2024 to March 2025. Each FPGA channel post lock-in was sampled at a frequency of 238 Hz, resulting in continuous phase and quadrature $Y(t)$ data streams, which were then broken into segments consisting of $20\times2^{14}$ samples each, or about 23 minutes of data. These output voltage streams can be referred back to the crystal displacement by applying the transfer function
	\begin{equation}\label{eq:strain}
		x(t) = \mathcal{F}^{-1}\left(\mathcal{F}\left(\frac{\sqrt{X(t)^2+Y(t)^2}}{2 \pi f_\lambda \kappa_\lambda G_s}\right)\right).
	\end{equation}
	Where $G_s$ denotes the signal gain between the SQUID input coil and the FPGA input, and the constant $\kappa_\lambda$ with units of Cm$^{-1}$ parameterises the piezoelectric coupling between the acoustic crystal and its electrodes of the overtone mode $\lambda$ with resonant frequency $f_\lambda$ \cite{Goryachev2014b}. In computing eq. (\ref{eq:strain}), we have utilized eq. (2) of ref. \cite{Campbell2023}, as well as the Fourier transform $\mathcal{F}$ and it's inverse. Following the calibration procedures of ref. \cite{Campbell2023}, the source independent strain sensitivity on resonance can be determined in units of $\mathrm{[strain]}/\sqrt{\mathrm{Hz}}$, and is shown for each mode in figure \ref{fig:setup} b).
	
	From Eq. (\ref{eq:strain}), $x(t)$ gives the instantaneous vibrational amplitude of the center of mass for the normal mode $\lambda$. A useful metric will be the energy of a normal mode which can be expressed as $k_b T(t)= \frac{1}{2}m_\lambda \dot{x}(t)^2$, where $k_b$ is Boltzman's constant, $m_\lambda$ is the normal mode effective mass and the dot denotes the derivative with respect to $t$. We can thus express the energy of a normal mode as an effective temperature $T(t)$ in units of Kelvin.
	
	Following established techniques from previous gravitational wave detection experiments~\cite{Allen2012}, we employ a matched filtering approach to search the detector output for signatures of HFGWs. When targeting transient signals $h(t)$ buried beneath a stationary noise process $n(t)$ in some detector output stream $x(t) = h(t) + n(t)$, matched filtering gives the following optimal signal to noise ratio (SNR) $\rho$, where $\tilde{h}(f)$ is the Fourier transform of the expected signal $h(t)$ and $S_n(f)$ is the single sided spectral density of the noise \cite{Tobar1995,Maggiore2008},
	\begin{equation}\label{eq:of}
		\rho^2 = 4\int_{0}^{\infty}df\frac{|\tilde{h}(f)|^2}{S_n(f)}.
	\end{equation}
	. 
	
	Applying the matched filter to each MAGE data segment, the spectral density of the detector noise $S_n(f)$ is calculated for each segment such that the matched filter is the adaptive Weiner-Kolmogorov function \cite{Heng1999}. This allows for a robust filter that adapts to time periods of increased detector noise.
	
	We aim to select collections of samples or `triggers where $\rho$ exceeds some threshold level corresponding to a transient increase in the vibrational amplitude of the crystal detector. Thus, the expected signal shape $h(t)$ is that of a decaying exponential $h(t) = h_0e^{-t/2\tau_\lambda}$ where $\tau_\lambda$ is the response time of the crystal mode. Due to the extraordinarily high quality factors of quartz BAW resonators, $\tau_\lambda$ is usually on the order of a second. Transient energy impulses inherent to the crystal can therefore be distinguished from other non-Gaussian noise sources by implementing a template bank with multiple values of decay time $\tau_b = \{\tau_1,...,\tau_\lambda,...,\tau_i\}$ and only selecting the candidate triggers for which $\rho$ is optimised for $\tau \sim \tau_\lambda$.
	
\begin{figure}
		\centering
		\includegraphics[width=0.5\textwidth]{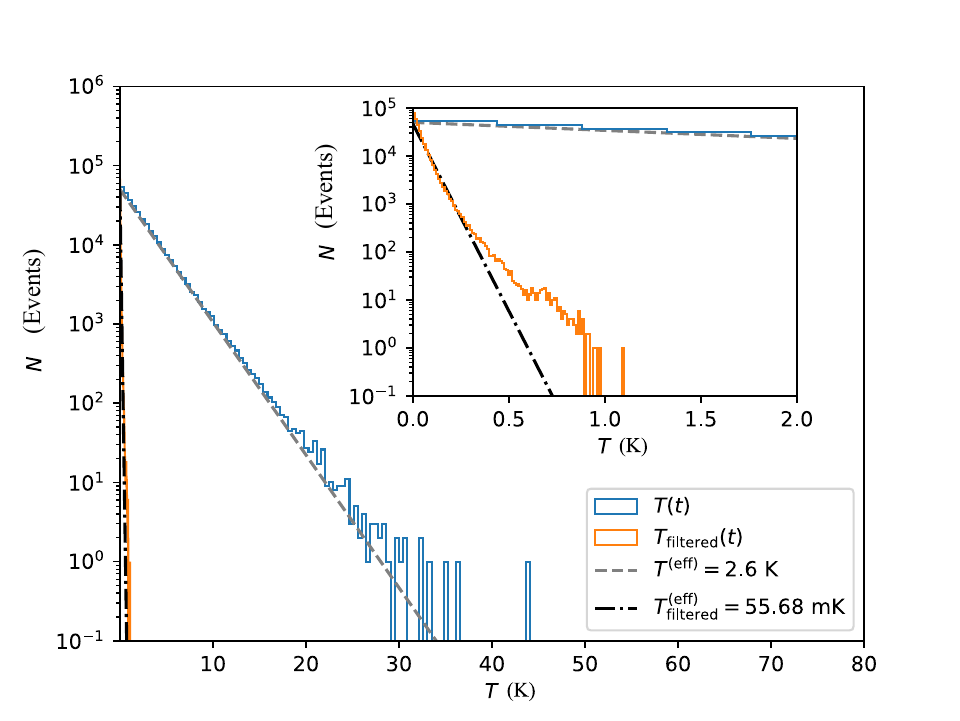}
		\caption{\label{fig:hist1}Instantaneous vibrational energy of a single mode is plotted as a histogram for a 23 minute segment of data. The blue (orange) histogram represents the energy distribution before (after) optimal filtering. Both histograms clearly follow an expected $\chi^2$ distribution, however the effect of optimal filtering greatly reduces the effective temperature at which events can be identified with SNR = 1.}
	\end{figure}

	Applying the optimal filter gives an SNR time series $\rho(t)$ which has unity mean and represents the signal-to-noise ratio of excess narrowband fluctuations above the thermal Nyquist noise limit of the crystal. To further understand the effects of optimal filtering we plot the energy histograms for a typical segment of data both before filtering ($T(t))$ and after ($T_\mathrm{filtering}(t)$) in Fig. \ref{fig:hist1}. Both histograms follow an exponential scaling attributed to a $\chi^2$ distribution, suggesting that the detector is dominated by Gaussian noise. This scaling follows $N\propto e^{(-T/T^{(\mathrm{eff})})}$, where $T^{(\mathrm{eff})}$ represents an effective `noise temperature' at which SNR = 1. It is thus a measure of the detectors' sensitivity to transient responses.
	
	The post-filtering energy distribution gives an effective noise temperature $T^{(\mathrm{eff})}_\mathrm{filtered}$ much lower than that of the physical mode temperature $\langle T_\lambda\rangle $, a common result for resonant mass gravitational wave detectors \cite{Astone1994}. As the optimal filter is constructed with knowledge of the averaged thermal noise spectrum $S_n(f)$,  $T_\mathrm{filtered}(t)$ represents the narrow band fluctuations inherent to the mode that are in excess of its equilibrium temperature $\langle T_\lambda\rangle$. 
	
	Observing Fig. \ref{fig:hist1}, we can see the presence of high-energy non-Gaussian fluctuations in the distribution tail, only observable after filtering. We also note that $T^{(\mathrm{eff})}\neq\langle T_\lambda\rangle$, this is due to white noise contributions to $S_n(f)$ from outside of the detector bandwidth. However, this noise is filtered out by the optimal filter and does not affect further results.
	
	Triggers are selected by applying the optimal filter to each segment of data, and selecting the local maxima $\rho > \rho_t$, where $\rho_t=1$ corresponds to an excitation of energy $T = T^{(\mathrm{eff})}_\mathrm{filtered}$. These maxima are then passed through the filter bank $\tau_b$ and refined such that the only remaining triggers are those that maximise $\rho$ for $\tau=\tau_\lambda$.
	
	\begin{figure}
		\centering
		\includegraphics[width=0.5\textwidth]{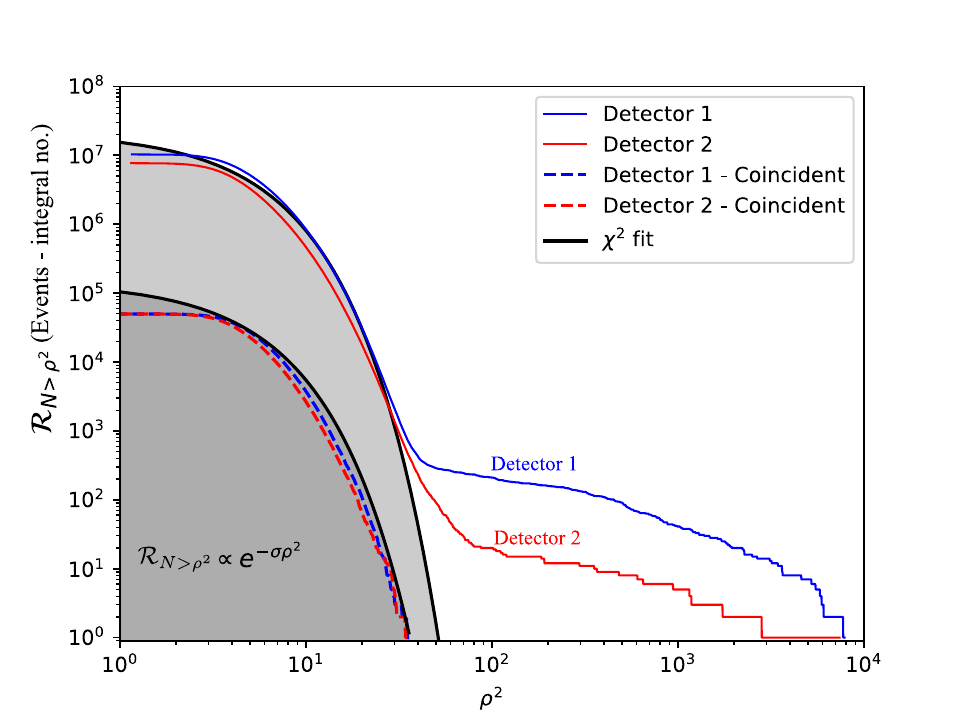}
		\caption{\label{fig:hist2}Integral histogram of triggers that passed the selection criteria $\mathcal{R}_{N >\rho^2}$ are shown as a function of energy SNR $\rho^2$. Dashed lines indicate the triggers that remain after the data set is refined by coincident counts. A $\chi^2$ model fit to the distributions of detector one is shown by the black lines. Clear separation between two distributions can be seen be defined by the shaded region under the $\chi^2$ fit.}
	\end{figure}
	The integral rate of triggers $\mathcal{R}_{N >\rho^2}$ for all modes across both detectors over 61 days of data collection is presented in the cumulative histogram in Fig. \ref{fig:hist2}. As energy is proportional to the square of $\rho$, the intrinsic thermal distribution of the crystal can be observed at low energies. However a large tail of rare high-energy triggers is also observed. These high-energy triggers are consistent with the significant events reported on in previous iterations of this experiment \cite{Goryachev2021}. The signal shape, significant statistical amplitude and rate, and simultaneous excitation in multiple overtone modes of the same crystal, are features common to all observations.
	
	Fitting a $\chi^2$ model $\mathcal{R}_{N>\rho^2}\propto e^{-\sigma\rho^2}$ for a scale constant $\sigma$, one can easily denote the separation between inherent thermal fluctuations of the crystal, and high-energy excitations in the distribution tail. A fitted model is plotted in Fig. \ref{fig:hist2} for to the triggers of detector one, and thus the thermal distribution can be observed as the shaded region. Deviations from the fit for $\rho \lesssim 3$ arise due to under-sampling of the thermal distribution, as the trigger selection algorithm only searches for transient local maxima in $\rho(t)$.
	
	To distinguish between gravitational signals and unwanted noise processes, MAGE introduces coincident analysis between the two detectors. Unwanted triggers can thus be removed by only keeping those common to both detectors within the same sampling bin. Applying this procedure removes the vast majority of triggers shown in Fig. \ref{fig:hist2}. The coincident histogram is observed to closely follow the $\chi^2$ model for the same value of $\sigma$ as the total parent distribution. This is expected as $\sigma$ is proportional to the thermal temperature of the device. From these observations all high-energy background triggers in the distribution tail can then be excluded as non-gravitational due to their lack of coincident detection.  
	
	Light PBH mass mergers are hypothesised to generate rapidly evolving HFGW signatures in the MHz frequency band of MAGE. In particular a binary system of equal PBH mass $m_\mathrm{PBH} = 4.4\times 10^{-3} M_{\odot}$ will emit maximal HFGW radiation at $f=5$ MHz during its innermost stable circular orbit (ISCO), although they rapidly pass through the narrow resonant band of a single overtone mode on a timescale as small as 10 ps \cite{Muia2022} . If such a binary system exists at a close enough distance, it will generate a coincident trigger in the MAGE system with an SNR corresponding to the signals strain amplitude. At such short time scales the signal can be treated as an impulse that excites the resonant modes of our detector, leading to an exponential decay of the vibrational amplitude on the inherent timescale of the detector mode. If the modes quality factor is large enough, than this amplitude will be detectable.
	
	Considering the distribution of coincident triggers and finding the maximum $\rho$ for each overtone mode $\lambda$ that gives at least $\rho\ge3$ in both detectors 1 and 2, an excluded confidence bound on gravitational wave strain can be determined. As the threshold level to detect a trigger is $\rho_t=1$, a potential HFGW signal with SNR $\bar{\rho} = 3$ would result in a confidence limit $\int_0^{\bar{\rho}} P(\rho|\rho_t) = 97.7\% $ \cite{Maggiore2008}.
	
	A useful metric to characterise the strength of merger signals with frequency evolution is the characteristic strain $h_c$. For an excluded coincident SNR $\bar{\rho}_\lambda$ that corresponds to a displacement amplitude ${x}_{\lambda}$ of the overtone mode $\lambda$, the minimum detectable characteristic strain at resonant frequency $f_\lambda$ is given by \cite{Campbell2023, Aggarwal2021, Aggarwal2025} 
	\begin{equation}\label{eq:hcdet}
		h_{c, \lambda} > {x}_{\mathrm{max},\lambda} \left|\frac{-2\pi^2 f^2 L_z\xi}{(2i \pi f)^2+\tau_\lambda^{-1}+(2\pi f_\lambda)^2} \right|^{-1}\frac{f_\lambda}{\Delta {f_\lambda}}.
	\end{equation}
	\begin{figure}
		\centering
		\includegraphics[width=0.5\textwidth]{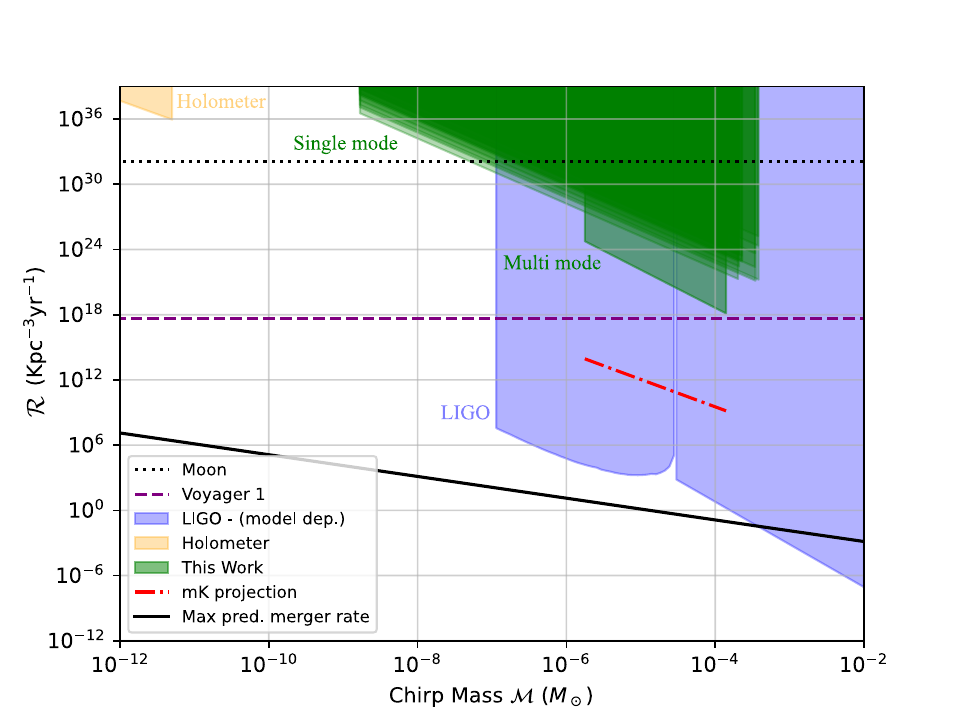}
		\caption{\label{fig:PBH}The excluded merger rate density of PBH binary systems determined in this work is plotted as the green shaded region. Also shaded are areas that have been excluded by previous experiments \cite{Chou2017,Miller2024,Miller2022}, although it must be noted that the LIGO exclusions rely on certain cosmological assumptions that are not made in this work \cite{Miller2024}. Dashed lines give the density that excludes one event per year at a distance corresponding to various astrophysical objects for reference. The solid black line is the best case possible merger rate density of PBH binaries if they constituted 100\% of local dark matter \cite{Muia2022}. The red line represents an idealistic estimate for a future version of MAGE operating at a reduced temperature of 10 mK.}
	\end{figure}
	Where $L_z$ is the detector crystal thickness, the sensitive bandwidth is denoted as  $\Delta f_\lambda$ \cite{Aggarwal2025}, and $\xi$ is a coupling term that parameterises the geometric overlap between gravitational and crystal acoustic fields \cite{Goryachev2014b}. The middle term in Eq. (\ref{eq:hcdet}) is the transfer function of the mechanical resonator, and the last term converts the dimensionless strain into detector characteristic strain by accounting for the sensitivity to frequency evolving in-spiral signals passing through the detectors narrow band. As $\Delta f_\lambda$ is usually of the order $<1$ Hz, this term is of order $\mathcal{O}(10^{7})$, highlighting the importance of broadband sensitivity when looking for rapidly evolving HFGW signals from PBH mergers.
	
	The characteristic strain excluded by the detector can be related to the strain amplitude $h_0$ of a PBH in-spiral of binary chirp mass $\mathcal{M}$ by 
	\begin{equation}
		h_c^2 = (2f \tilde{h}(f))^2 = 2 h_0^2 N_\mathrm{cycles}(\mathcal{M}, f),
	\end{equation} 
	Where $N_\mathrm{cycles}(\mathcal{M},f)$ is the number of cycles the in-spiral signal spends at the frequency $f$, its explicit form can be found elsewhere \cite{Moore2015}. Eq. (\ref{eq:hcdet}) can thus be related to the maximum distance of reach of the detector $d_\mathrm{max}$ by considering the post-Newtonian approximation for an in-spiralling source at distance $d$, emitting gravitational radiation at a frequency $f$ \cite{Antelis2018}.
	\begin{equation}
		h_0\approx\frac{2}{d}\left(\frac{G\mathcal{M}}{c^2}\right)^{5/3}\left(\frac{\pi f}{c}\right)^{2/3}
	\end{equation}
	Where $G$ is Newtons gravitational constant. For each mode $\lambda$ we then obtain through coincident analysis of detector 1 and 2  a resonant bound on the maximum distance of reach $d_{\mathrm{max}, \lambda}$ for PBH binaries of equal mass such that the ISCO frequency is equal to $f_\lambda$. However one can extend this bound to other PBH masses in which the in-spiralling signal will pass through the detector band on its way to an ISCO frequency $f_\mathrm{ISCO}>f_\lambda$. The lower mass limit to this approach is given from constraining the signal lifetime as it passes through the detector band to be less than the sampling time such that it registers as a single-bin coincident trigger in $\rho(t)$.
	
	A more optimised approach to excluding PBH binaries can be achieved by exploiting the multiple overtone modes of MAGE to search for an in-spiral signal that passes through each and every detector mode $\lambda$. For a detector with sensitivity in $N_\lambda$ discontinuous narrow frequency bands, Eq. (\ref{eq:of}) collapses to
	\begin{equation}
		\rho ^2 = 2\int_{0}^{\infty}df\frac{N_\mathrm{cycles}(f) h_0^2}{f^2S_n(f)} \sim \frac{h_0^2}{2}\sum_\lambda^{N_\lambda} \frac{\Delta f_\lambda^2~ N_\mathrm{cycles}(f_\lambda)}{f_\lambda^2~h_n^2(f_\lambda)}
	\end{equation}
	The strain amplitude that would generate an SNR of $\rho$ is thus given by
	\begin{equation}\label{eq:h-sum}
		h_0^2 \sim 2\rho^2 \left[\sum_{\lambda=1}^{N_\lambda}\frac{\Delta f_\lambda^2~N_\mathrm{cycles}(f_\lambda)}{f_\lambda^2 h_n^2(f_\lambda)}\right]^{-1},
	\end{equation}
	and thus for each overtone mode up to $ N_\lambda$ the detector gains effective bandwidth, and sensitivity to characteristic PBH strain increases.
	
	As no triggers are observed to be coincident in all 32 detector modes, a distance of reach can be excluded to 97.7\% confidence during the observational period by determining the inverse sum of (\ref{eq:h-sum}) excluding a signal of SNR $\bar{\rho}=3$. For each mode, the strain sensitivity on resonance $h_n(f_\lambda)$ is determined by finding the mean amplitude of triggers that gives $\rho=\bar{\rho}$. This is equivalent to finding the strain that corresponds to a vibrational energy $\bar{\rho}^2~T^{\mathrm{(eff)}}_\mathrm{filtered}$.
	
	Eq. (\ref{eq:hcdet}) and (\ref{eq:h-sum}) both give bounds on the maximum distance of reach to which a PBH merger would be observable through single-mode, and multi-mode analysis methods, respectively. Considering the total observation time of $ T=$61 days, these bounds can be converted into the convenient units of merger rate density $\mathcal{R} = ( \frac{4}{3}\pi d_\mathrm{max}^3T)^{-1}$. The resulting PBH merger rate limits are presented in Fig. \ref{fig:PBH}, where PBH binaries are excluded to some level for binary masses $1.2\times10^{-4}M_\odot<\mathcal{M}<1.7\times10^{-9}M_\odot$. The multi-mode approach gives the strongest bound of $\mathcal{R}<1.3\times10^{18}~\mathrm{kpc}^{-3}\mathrm{yr}^{-1}$, corresponding to a distance of reach $d=1.0\times10^{-6}$ kpc. Whilst it can be observed that interferometers have excluded a large range of parameter space \cite{Miller2022,Miller2024} these limits crucially depend on the PBH signal being monochromatic in the early stages of its in-spiral. These constraints come into question in scenarios that feature perturbations to the ideal circular two body orbit of PBHs such as large eccentricity or dark matter clustering.  The limits provided in this work thus represent a complimentary approach with direct and model-independent sensitivity to PBH in-spirals in the high frequency regime.
	
	From Fig. \ref{fig:PBH}, the sensitivity of MAGE is observed to be orders of magnitude away from a distance of reach which could detect a PBH binary as some population of dark matter. In order to reach these bounds one would require $\mathcal{O}(10^{6})$ gain in sensitivity to characteristic strains. In the supplemental material \footnote{See supplemental material for the detailed sensitivty analysis of an acoustic HFGW detector dominated by 10 mK thermal noise. Includes Refs. \cite{Galliou2013, Astone1992}} we model the sensitivity of MAGE at the extreme cryogenic temperature of 10mK. At these lower temperatures the gains in sensitivity for MAGE could see a large improvement to extend its distance of reach well beyond the vicinity of the solar system. Considering appropriate increases to $Q$ and quantum limited amplification one can estimate a maximum distance of reach of $5\times10^{-4}$ kpc, corresponding to the merger density rate observed as the red dashed line in Fig. \ref{fig:PBH}. However, further technological advancements are still necessary to reach the levels of sensitivity required to constrain PBH mergers as a fraction of local dark matter. 	
	
	Intrinsic quality factors in acoustic crystals at cryogenic temperatures are known to be limited by defects and impurity sites \cite{Galliou2013, Goryachev2013}, improvements to fabrication techniques have recently lead to record quality factors for BAW resonators on the nano-mechanical scale \cite{Kharel2018, Luo2025} by mitigating such defects. Adapting these techniques to larger scales could see similar improvements to the MAGE detectors. Low temperature properties of other suitable piezoelectric materials with larger densities than quartz are also currently under investigation \cite{Campbell2025}. Increased effective masses will lead to stronger acoustic-gravitational couplings and thus further sensitivity. The multi-mode nature of MAGE can also be leveraged by additional detectors and overtone modes in an array. Thanks to the small form factor of the MAGE detector \cite{Campbell2023} packing multiple crystals into a single cryogenic systems is achievable. Additionally, rapid scanning approaches could see further sensitivity to in-spiral signals \cite{Campbell2023b} if the tuning can be implemented on the short time scales of in-spiral evolution. Whilst it is clear that the detection of HFGWs remains technologically extremely challenging, MAGE represents a significant step forward for dedicated experimental approaches.
	
	This research was supported by the ARC Centre of Excellence for Engineered Quantum Systems (CE170100009) and the ARC Centre of Excellence for Dark Matter Particle Physics (CE200100008).
	\bibliography{PBH}
\end{document}